\documentclass[sigconf]{acmart}

\usepackage{booktabs}
\usepackage{cleveref}
\usepackage{dirtytalk}
\usepackage{natbib}
\usepackage{subfig}
\usepackage{enumitem}
\usepackage{url}
\usepackage{bbding} 
\usepackage{framed}
\usepackage{lscape}

\copyrightyear{2018}
\acmYear{2018}
\setcopyright{acmlicensed}
\acmConference[ESEM '18]{ACM / IEEE International Symposium on Empirical Software Engineering and Measurement (ESEM)}{October 11--12, 2018}{Oulu, Finland}
\acmPrice{15.00}
\acmDOI{10.1145/3239235.3267437}
\acmISBN{978-1-4503-5823-1/18/10}

\begin{document}

\title[Standards of Validity and the Validity of Standards]{Standards of Validity and the Validity of Standards in \\Behavioral Software Engineering Research: \\The Perspective of Psychological Test Theory}

\author{Lucas Gren}
%\authornote{Dr.~Trovato insisted his name be first.}
\orcid{1234-5678-9012}
\affiliation{%
  \institution{Chalmers University of Technology and The University of Gothenburg\\}
  \streetaddress{The Department of Computer Science and Engineering}
  \city{Gothenburg} 
  \state{Sweden} 
  \postcode{412--92}
}
\email{lucas.gren@cse.gu.se}

\begin{abstract}
\paragraph{Background}
There are some publications in software engineering research that aim at guiding researchers in assessing validity threats to their studies. Still, many researchers fail to address many aspects of validity that are essential to quantitative research on human factors.

\paragraph{Goal}
This paper has the goal of triggering a change of mindset in what types of studies are the most valuable to the behavioral software engineering field, and also provide more details of what construct validity is.

\paragraph{Method}
The approach is based on psychological test theory and draws upon methods used in psychology in relation to construct validity. 

\paragraph{Results}
In this paper, I suggest a different approach to validity threats than what is commonplace in behavioral software engineering research.

\paragraph{Conclusions}
While this paper focuses on behavioral software engineering, I believe other types of software engineering research might also benefit from an increased focus on construct validity.

\end{abstract}


%\ccsdesc[300]{Applied computing~Psychology}

%
% End generated code
%

%
%  Use this command to print the description
%

% We no longer use \terms command
%\terms{Theory}


\settopmatter{printfolios=true}
\maketitle

\section{Introduction}
Most reviewers and authors in the behavioral software engineering research field, base their concept of validity on two publications, namely \citet{expbookwholin} and \citet{runeson}. While I highly appreciate the authors' work in these publications, I am afraid the four categories of validity threats promoted for both experimentation and case study research (i.e.\ construct, internal, external, and reliability), have often turned into a \emph{quick fix} at the end of papers. This tendency might be because the categories are not described in more detail concerning test theory. In this paper I will firstly go through this problem in more detail and then suggest how psychological test theory can be used to overcome parts of the problem. 

\section{The Problem}
A stepwise and shallow presentation of the four validity threats categories, at the end of each paper, indicates an immature approach to validity aspects in the behavioral software engineering field, even for well-designed studies regarding the resources available and novelty of the work presented. Besides, I have met very few researchers who remember the name of the first category and what it comprises, i.e.\ construct validity, simply because researchers have read other categories in related fields, such as criterion, conclusion, concurrent, content, and convergent validity. Sadly, we also lose many important subcategories of construct validity that should be taken into account throughout any research project or program, especially the ones promoting validation with real data. Such studies are of course demanding and require a lot of resources, which imply that they take time.

In accordance with \citet{feldt2010validity}, I believe that ``validity is a goal, not something that can be proven or assured with the use of specific procedures.'' There are also trade-offs between different validity threats, e.g.\ if more resources are spent in assuring a good operationalization of a particular cause and effect, the whole experiment might suffer from being a \emph{toy problem} with very little or no practical value. Therefore, I do not agree that as many validity threats as possible should be listed at the end of each publication together with statements of how they were mitigated, since such a Utopian study does not, and will never, exist. Instead, I suggest that the \emph{method} section should be reworked until it is as clear as can be, in line with how \citet{jedlitschka2005reporting} suggest controlled experiments should be designed, however, these do not have to be explicitly stated as validity threats since that is obvious. With a well-written method section, threats to validity will be clear in the description of the planning conducted, and statements regarding sample sizes and generalization need not be explicitly stated or repeated in a discussion of validity threats at the end of each paper. As \citet{feldt2010validity} also conclude, the guidelines given in both \citet{expbookwholin} and \citet{runeson} are from conducting experimental planning or case studies. They were not intended as post-study \emph{quick fix} checklists by the authors, which they also explicitly state in \citet{runeson}: ``It is, as described above, important to consider the validity of the case study from the beginning.''

\subsection{Psychological reasons for poor validity}
I believe a systematic approach to threats to validity is useful in research, however, there is a difference between having a checklist for a research program and including four categories of threats in the end of a paper. A checklist with as many categories as possible should be used, I argue, to better address biases. 

Some troublesome studies about research evaluations show that researchers, just like all people, often conduct a biased memory search based on their motivation. However, people are not biased without the feeling of being able to justify their conclusions. In a study by \citet{gilovich1983biased} the loss of a football team was explained by a fluke in the game only if the subjects were aware of it, and if they were not, they lost faith in their team's talent. There is also evidence showing that directional goals may affect the use of statistical heuristics. In a study by \citet{ginossar1987problem}, they showed that subjects used information (like a base-rate) only when it could motivate reaching their directional goal. By re-analyzing such results, \citet{kunda1990case} proposes that the infamous results by the Nobel Prize winners of economic sciences, Tversky and Kahneman popularized by \citet{kahneman2011tfa}, is an oversimplification and that the more analytical reasoning is also prone to bias. \citet{kunda1990case} also shows that the more generally known confirmation bias \citep{nickerson1998confirmation}, is not purely a cognitive bias, but better understood as a biased memory search (even if Kunda herself did not use the term \emph{confirmation bias}). In a troublesome review of biased research evaluations, she presents cases where subjects in favor of a specific directional goal judged research studies as of higher validity and better conducted as compared to subjects with a different predisposition (see, e.g.\ \citet{pyszczynski1987toward}). She concludes that the subjects used heuristics depending on the conclusions of the research, not its methods \citep{kunda1990case}, which is, again, utterly troublesome for the academic community. Since this is the case, we need an utterly structured approach at the other end, i.e.\ to demand a structured and in depth approach to validity in the research designs.  The question is if we will ever be able to entirely manage confirmation bias in research. At least, we can be structured and consistent in our requirements of the different scientific methods used.

The confirmation bias problem in research has also been shown to exist in software engineering explicitly \citep{jorgensen2015believing}, which is far from surprising, but important to show. On top of confirmation bias comes both researcher bias (statistically non-significant results that become significant through questionable research or analysis practices) and publication bias (statistically non-significant results that are not reported), something \citet{jorgensen2016incorrect} also have studied in the empirical software engineering domain. All these studies show that we need a more systematic approach to validity on many different levels.

Generalizations need to be drawn with more care, I argue, and even the most extensive empirical studies in software engineering have a too small sample size to state anything about the \emph{truth} for these concepts. It takes a field decades to build up a body of knowledge extensive enough for a meta-study to have such claims of external validity. See for example \citet{freemanpnas}, where they used 225 studies to conclude that active learning outperforms traditional lecturing with regards to student performance. Or the conclusion that we inherit 49\% of our developed human traits, based on 2,748 publications including 14,558,903 twin pairs \citep{polderman2015meta}. The point is that new concepts in exploratory research are most likely not possible to generalize outside of the specific case. We could choose to believe that it is true somewhere else, but without empirical evidence to support such a claim. However, the internal validity is often considered higher in interview studies since a validation of the possible causal relationships is included in the design. 

Validity of research is a thorny issue and of course depend on the research design, however, I believe a larger focus on construct validity is needed both in behavioral software engineering and, parts of what I suggest below, are also applicable to more general software engineering studies. 

\section{The Solution}\label{solution}
I suggest that no categories be used for listing validity threats in research papers since I believe they are somewhat counterproductive, despite the fact that I have most often done that myself. The more complex validity aspects that need clarification after the reader has read the method section could be discussed under a section called \emph{Validity threats}, \emph{Threats to validity}, \emph{Limitations}, or the like. Another option is to simply write the threats as a part of the discussion section, but the practical significance of the threats needs to be in focus. What is important is that researchers consider threats to validity throughout their research. To help to implement such an awareness, I suggest a checklist below inspired by seminal work conducted by \citet{messick} in relation to testing validity, but where I also include validity aspects in relation to specific research studies as already suggested by \citet{expbookwholin} and \citet{runeson}.

\citet{messick1995standards} defines validity as follows: 
\begin{quote}
``Validity is not a property of the test or assessment as such, but rather of the meaning of the test scores. These scores are a function not only of the items or stimulus conditions but also of the persons responding as well as the context of the assessment. In particular, what needs to be valid is the meaning or interpretation of the score; as well as any implications for action that this meaning entails.''
\end{quote}
This definition implies that we always validate the usage of a test, and never the test itself. I would like to mention here that a \emph{test} refers to a psychological test, i.e.\ a measurement of a construct (a \emph{construct} in any scientific field is a phenomenon defined as a distinct category that is under study). In the quote by Messick, we can see that he advocates a more applied and practical treatment of validity. He also argues that validity of a test is only one construct that he calls \emph{construct validity}. He writes that different aspects of construct validity can still be presented in order of convenience. However, they are still interrelated both operationally and logically. Also ``the principles of validity apply to all assessments, whether based on tests, questionnaires, behavioral observations, work samples, or whatever'' \citep{messick1995standards}. The presented six aspects are Consequential, Content, Substantive, Structural, External, and Generalizability, and, to clarify, are all concerning the actual measurements, and not, for example, the generalizability of treatment in a specific experiment.

In addition, reliability is seen as a prerequisite for validity, and the external, internal and conclusion validity in relation to a research study is also included below, following \citet{expbookwholin} and \citet{runeson}. However, I think it is of utter importance to change the culture in software engineering research from seeing tool-constructing as the holy grail of research and instead value validation studies higher, which then also includes validation of existing tools. To build theory, and make good use of research funding, software engineering researchers need to conduct; ideally, a study of each aspect of construct validity presented below, before drawing conclusions to the intended population and connections between constructs. As already stated, my example is from behavioral software engineering research, but are partly applicable to closely related sub-fields of software engineering. Throughout a research project or program, I suggest the following checklist be used: 

\begin{itemize}
\item Reliability --- Is \emph{repeatability} or \emph{consistency}. A measure is considered reliable if it would give us the same result over and over again (assuming that what we are measuring is static), essentially answering the question: Does the test measure anything? 
    \begin{enumerate}
        \item Stability --- Is the testing stable if we do a test and then another test on the same subjects under the same conditions (test-retest procedure) or parallel testing?
        \item Internal consistency --- Is the test consistent with regards to, e.g. the Kuder-Richardson Formula 20 (KR20) \citep{kuder1937theory} or Cronbach's $\alpha$ \citep{cronbach}?
    \end{enumerate}

\item Construct Validity --- Does the test measure what it is meant to measure? 
    \begin{enumerate}
        \item Consequential --- What are the potential risks if the scores are, in actuality, invalid or inappropriately interpreted? 
        \item Content --- Do test items appear to be measuring the construct of interest?
        \item Substantive --- Is the theoretical foundation underlying the construct of interest sound?
        \item Structural --- Do the interrelationships of dimensions measured by the test correlate with the construct of interest and test scores? These can be tested by using a factor analysis (FA) or a principal component analysis (PCA). 
        \item External --- Does the test have ecological, convergent, discriminant, and predictive qualities?
                \begin{enumerate}
                    \item Ecological --- Is the real-world behavior in accordance with how a subject answers the test?  
                    \item Convergent --- Is the test similar to (converges on) other operationalizations that it theoretically should be similar to? For example, two agile maturity models should result in similar levels of maturity when applied to the same organization at the same time. 
                    \item Discriminant ---  Is the test dissimilar to (diverges from) other operationalizations that it theoretically should not be similar to? Maybe we hypothesize that testing practices should not be correlated to developers' self-esteem. If so, measurements of both should have a low correlation.
                \item Predictive --- Can the tests predict something it should theoretically be able to predict?
    \end{enumerate}
        \item Generalizability --- Does the test generalize across different groups, settings, and tasks? This aspect of generalizability is in relation to the actual measurement only, i.e.\ generalizability of one measured construct only. This category is not to be confused with external validity of the whole study described below.  
    \end{enumerate}

\item Conclusion validity ---  Is the degree to which conclusions reached regarding relationships in our data reasonable? 

\item Internal validity ---  Internal validity is most often seen as an investigation of causality between measured constructs, i.e.\ could there be other confounding factors in the study where the test(s) is(are) used?

\item External validity ---  External validity is concerned with generalizations in relation to the \emph{whole} study. Are the results valid for the intended larger population? In the case of experimentation, this includes aspects like treatment, subjects, and context. In the case of a correlative survey study, the external validity would be the generalizability of the actual correlation study and not the constructs being measured. In the ideal case, a smaller correlation study deploys measurements that have already been validated with large datasets. 

\end{itemize}
For the sake of completeness in regard to validity threats categories sometimes mentioned in research, criterion validity is sometimes used to describe predictive validity (as stated above) and a category called \emph{concurrent} validity \citep{American2014sfe}. I have chosen to exclude that category above and the parent category of criterion validity since concurrent validity refers to the test correlated against a benchmark. In empirical software engineering research the first step will be to create valid measurements and we still do not have highly valid benchmark tests, in my opinion. However, this category might be useful sometime in the future but not for some years to come. Another common category excluded above is \emph{face} validity, which is simply an overall sanity check of the whole research project, i.e.\ is the project reasonable using common sense? 

\begin{figure*}
\centerline{\includegraphics[scale=0.7]{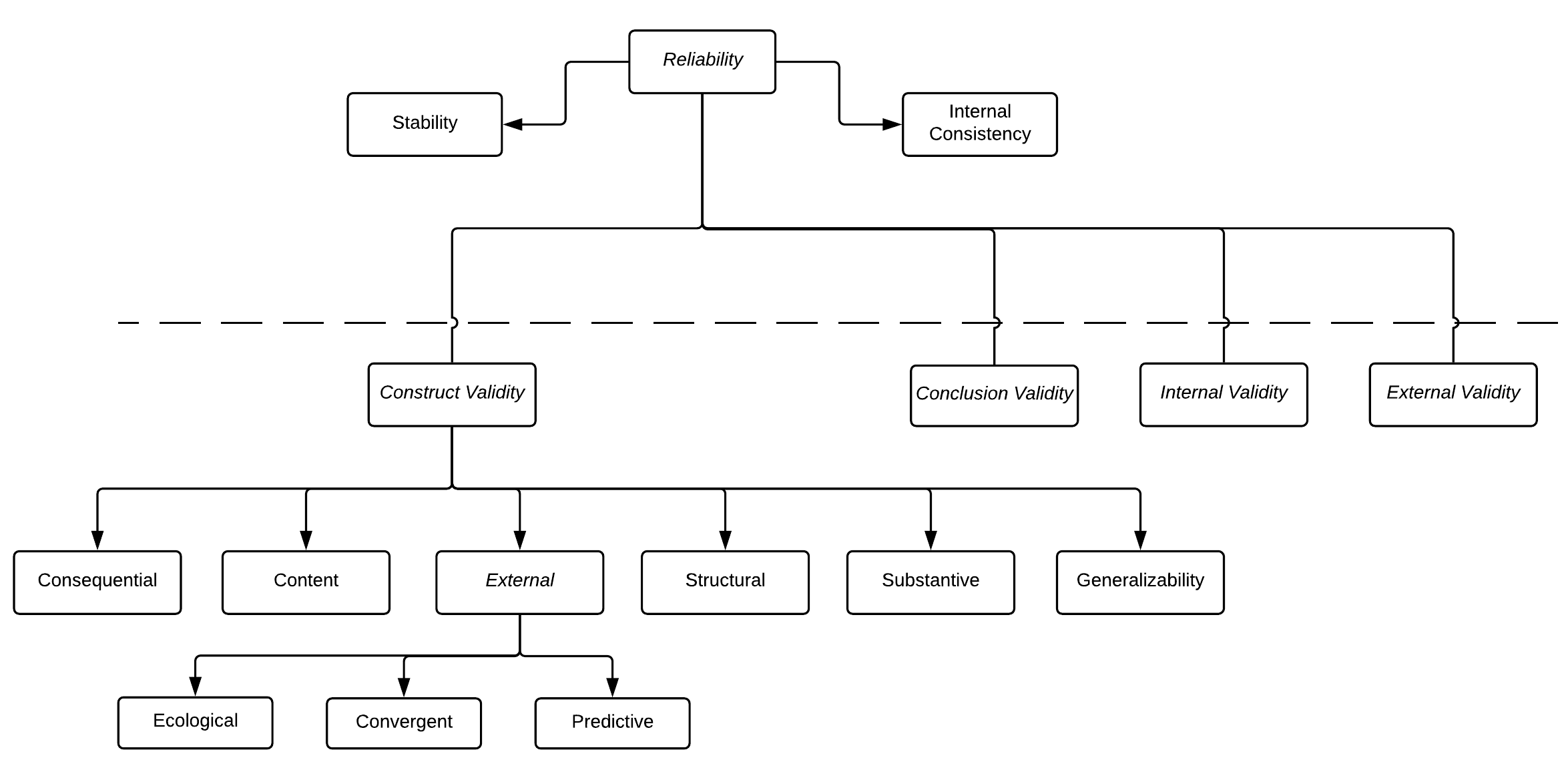}}
\caption{Hierarchical structure of the proposed checklist.}
\label{chart}
\end{figure*}

In order to obtain a clearer idea of the structure of the included categories, see Figure~\ref{chart}. 

%\begin{landscape}
%\end{landscape}

For examples of how to apply the checklist, please check \citet{grenphd}.

\section{Conclusions and Future Work}
On a final remark, I would like to state that analyzing and navigating through validity threats throughout a research program is a very challenging endeavor, as shown in the psychological discourse \citep{American2014sfe}. I think that my proposed integrated checklist is useful and could make one realize what validation studies are needed to be conducted in the future to strengthen the validity of the conclusions reached. As also mentioned, I think it is imperative to change the culture in software engineering research from seeing tool-constructing as the holy grail of research and instead value validation studies higher. This is especially true for behavioral software engineering research. Such studies are crucial to theory building and, just like in psychological research, spending an entire Ph.D.\ candidacy on the validation of one single measurement of a construct should be, not only approved, but encouraged. 

Now, I just pray my proposed integrated checklist will not be used as categories in the end of papers with short mitigation strategies in connection to them. But even so, including more categories would still be an improvement from what is conducted today.

%\begin{acks}
%I would like to thank Vard Antinyan for interesting discussions on this topic, possible explanations to the results, and feedback when writing this paper. 
%\end{acks}

\bibliographystyle{ACM-Reference-Format}
\bibliography{references}  

%%% -*-BibTeX-*-
%%% Do NOT edit. File created by BibTeX with style
%%% ACM-Reference-Format-Journals [18-Jan-2012].

\begin{thebibliography}{00}

%%% ====================================================================
%%% NOTE TO THE USER: you can override these defaults by providing
%%% customized versions of any of these macros before the \bibliography
%%% command.  Each of them MUST provide its own final punctuation,
%%% except for \shownote{}, \showDOI{}, and \showURL{}.  The latter two
%%% do not use final punctuation, in order to avoid confusing it with
%%% the Web address.
%%%
%%% To suppress output of a particular field, define its macro to expand
%%% to an empty string, or better, \unskip, like this:
%%%
%%% \newcommand{\showDOI}[1]{\unskip}   % LaTeX syntax
%%%
%%% \def \showDOI #1{\unskip}           % plain TeX syntax
%%%
%%% ====================================================================

\ifx \showCODEN    \undefined \def \showCODEN     #1{\unskip}     \fi
\ifx \showDOI      \undefined \def \showDOI       #1{{\tt DOI:}\penalty0{#1}\ }
  \fi
\ifx \showISBNx    \undefined \def \showISBNx     #1{\unskip}     \fi
\ifx \showISBNxiii \undefined \def \showISBNxiii  #1{\unskip}     \fi
\ifx \showISSN     \undefined \def \showISSN      #1{\unskip}     \fi
\ifx \showLCCN     \undefined \def \showLCCN      #1{\unskip}     \fi
\ifx \shownote     \undefined \def \shownote      #1{#1}          \fi
\ifx \showarticletitle \undefined \def \showarticletitle #1{#1}   \fi
\ifx \showURL      \undefined \def \showURL       #1{#1}          \fi
% The following commands are used for tagged output and should be
% invisible to TeX
\providecommand\bibfield[2]{#2}
\providecommand\bibinfo[2]{#2}
\providecommand\natexlab[1]{#1}
\providecommand\showeprint[2][]{arXiv:#2}

\bibitem[\protect\citeauthoryear{Association}{Association}{2014}]%
        {American2014sfe}
\bibfield{author}{\bibinfo{person}{American Educational~Research Association}.}
  \bibinfo{year}{2014}\natexlab{}.
\newblock \bibinfo{booktitle}{{\em Standards for educational and psychological
  testing}}.
\newblock \bibinfo{publisher}{American Educational Research Association},
  \bibinfo{address}{Washington, DC}.
\newblock


\bibitem[\protect\citeauthoryear{Cronbach}{Cronbach}{1951}]%
        {cronbach}
\bibfield{author}{\bibinfo{person}{Lee Cronbach}.}
  \bibinfo{year}{1951}\natexlab{}.
\newblock \showarticletitle{Coefficient alpha and the internal structure of
  tests}.
\newblock \bibinfo{journal}{{\em Psychometrika\/}} \bibinfo{volume}{16},
  \bibinfo{number}{3} (\bibinfo{year}{1951}), \bibinfo{pages}{297--334}.
\newblock


\bibitem[\protect\citeauthoryear{Feldt and Magazinius}{Feldt and
  Magazinius}{2010}]%
        {feldt2010validity}
\bibfield{author}{\bibinfo{person}{Robert Feldt} {and} \bibinfo{person}{Ana
  Magazinius}.} \bibinfo{year}{2010}\natexlab{}.
\newblock \showarticletitle{Validity Threats in Empirical Software Engineering
  Research -- An Initial Survey}. In \bibinfo{booktitle}{{\em International
  Conference on Software Engineering and Knowledge Engineering (SEKE)}}.
  \bibinfo{pages}{374--379}.
\newblock


\bibitem[\protect\citeauthoryear{Freeman, Eddy, McDonough, Smith, Okoroafor,
  Jordt, and Wenderoth}{Freeman et~al\mbox{.}}{2014}]%
        {freemanpnas}
\bibfield{author}{\bibinfo{person}{Scott Freeman}, \bibinfo{person}{Sarah~L.
  Eddy}, \bibinfo{person}{Miles McDonough}, \bibinfo{person}{Michelle~K.
  Smith}, \bibinfo{person}{Nnadozie Okoroafor}, \bibinfo{person}{Hannah Jordt},
  {and} \bibinfo{person}{Mary~Pat Wenderoth}.} \bibinfo{year}{2014}\natexlab{}.
\newblock \showarticletitle{Active learning increases student performance in
  science, engineering, and mathematics}.
\newblock \bibinfo{journal}{{\em Proceedings of the National Academy of
  Sciences\/}} \bibinfo{volume}{111}, \bibinfo{number}{23}
  (\bibinfo{year}{2014}), \bibinfo{pages}{8410--8415}.
\newblock
\showDOI{%
\url{http://dx.doi.org/10.1073/pnas.1319030111}}


\bibitem[\protect\citeauthoryear{Gilovich}{Gilovich}{1983}]%
        {gilovich1983biased}
\bibfield{author}{\bibinfo{person}{Thomas Gilovich}.}
  \bibinfo{year}{1983}\natexlab{}.
\newblock \showarticletitle{Biased evaluation and persistence in gambling}.
\newblock \bibinfo{journal}{{\em Journal of personality and social
  psychology\/}} \bibinfo{volume}{44}, \bibinfo{number}{6}
  (\bibinfo{year}{1983}), \bibinfo{pages}{1110}.
\newblock


\bibitem[\protect\citeauthoryear{Ginossar and Trope}{Ginossar and
  Trope}{1987}]%
        {ginossar1987problem}
\bibfield{author}{\bibinfo{person}{Zvi Ginossar} {and} \bibinfo{person}{Yaacov
  Trope}.} \bibinfo{year}{1987}\natexlab{}.
\newblock \showarticletitle{Problem solving in judgment under uncertainty}.
\newblock \bibinfo{journal}{{\em Journal of Personality and social
  Psychology\/}} \bibinfo{volume}{52}, \bibinfo{number}{3}
  (\bibinfo{year}{1987}), \bibinfo{pages}{464}.
\newblock


\bibitem[\protect\citeauthoryear{Gren}{Gren}{2017}]%
        {grenphd}
\bibfield{author}{\bibinfo{person}{Lucas Gren}.}
  \bibinfo{year}{2017}\natexlab{}.
\newblock {\em \bibinfo{title}{Psychological group processes when building
  agile software development teams}}.
\newblock \bibinfo{thesistype}{Ph.D. Dissertation}. \bibinfo{school}{The
  University of Gothenburg}.
\newblock


\bibitem[\protect\citeauthoryear{Jedlitschka and Pfahl}{Jedlitschka and
  Pfahl}{2005}]%
        {jedlitschka2005reporting}
\bibfield{author}{\bibinfo{person}{Andreas Jedlitschka} {and}
  \bibinfo{person}{Dietmar Pfahl}.} \bibinfo{year}{2005}\natexlab{}.
\newblock \showarticletitle{Reporting guidelines for controlled experiments in
  software engineering}. In \bibinfo{booktitle}{{\em International Symposium on
  Empirical Software Engineering}}. IEEE, \bibinfo{pages}{95--104}.
\newblock


\bibitem[\protect\citeauthoryear{J{\o}rgensen, Dyb{\aa}, Liest{\o}l, and
  Sj{\o}berg}{J{\o}rgensen et~al\mbox{.}}{2016}]%
        {jorgensen2016incorrect}
\bibfield{author}{\bibinfo{person}{Magne J{\o}rgensen}, \bibinfo{person}{Tore
  Dyb{\aa}}, \bibinfo{person}{Knut Liest{\o}l}, {and} \bibinfo{person}{Dag~IK
  Sj{\o}berg}.} \bibinfo{year}{2016}\natexlab{}.
\newblock \showarticletitle{Incorrect results in software engineering
  experiments: {H}ow to improve research practices}.
\newblock \bibinfo{journal}{{\em Journal of Systems and Software\/}}
  \bibinfo{volume}{116} (\bibinfo{year}{2016}), \bibinfo{pages}{133--145}.
\newblock


\bibitem[\protect\citeauthoryear{J{\o}rgensen and Papatheocharous}{J{\o}rgensen
  and Papatheocharous}{2015}]%
        {jorgensen2015believing}
\bibfield{author}{\bibinfo{person}{Magne J{\o}rgensen} {and}
  \bibinfo{person}{Efi Papatheocharous}.} \bibinfo{year}{2015}\natexlab{}.
\newblock \showarticletitle{Believing is seeing: {C}onfirmation bias studies in
  software engineering}. In \bibinfo{booktitle}{{\em 41st Euromicro Conference
  on Software Engineering and Advanced Applications (SEAA)}}. IEEE,
  \bibinfo{pages}{92--95}.
\newblock


\bibitem[\protect\citeauthoryear{Kahneman}{Kahneman}{2011}]%
        {kahneman2011tfa}
\bibfield{author}{\bibinfo{person}{Daniel Kahneman}.}
  \bibinfo{year}{2011}\natexlab{}.
\newblock \bibinfo{booktitle}{{\em Thinking, fast and slow}}.
\newblock \bibinfo{publisher}{Farrar, Straus and Giroux}, \bibinfo{address}{New
  York}.
\newblock


\bibitem[\protect\citeauthoryear{Kuder and Richardson}{Kuder and
  Richardson}{1937}]%
        {kuder1937theory}
\bibfield{author}{\bibinfo{person}{G~Frederic Kuder} {and}
  \bibinfo{person}{Marion~W Richardson}.} \bibinfo{year}{1937}\natexlab{}.
\newblock \showarticletitle{The theory of the estimation of test reliability}.
\newblock \bibinfo{journal}{{\em Psychometrika\/}} \bibinfo{volume}{2},
  \bibinfo{number}{3} (\bibinfo{year}{1937}), \bibinfo{pages}{151--160}.
\newblock


\bibitem[\protect\citeauthoryear{Kunda}{Kunda}{1990}]%
        {kunda1990case}
\bibfield{author}{\bibinfo{person}{Ziva Kunda}.}
  \bibinfo{year}{1990}\natexlab{}.
\newblock \showarticletitle{The case for motivated reasoning}.
\newblock \bibinfo{journal}{{\em Psychological bulletin\/}}
  \bibinfo{volume}{108}, \bibinfo{number}{3} (\bibinfo{year}{1990}),
  \bibinfo{pages}{480}.
\newblock


\bibitem[\protect\citeauthoryear{Messick}{Messick}{1995a}]%
        {messick1995standards}
\bibfield{author}{\bibinfo{person}{Samuel Messick}.}
  \bibinfo{year}{1995}\natexlab{a}.
\newblock \showarticletitle{Standards of validity and the validity of standards
  in performance assessment}.
\newblock \bibinfo{journal}{{\em Educational measurement: Issues and
  practice\/}} \bibinfo{volume}{14}, \bibinfo{number}{4}
  (\bibinfo{year}{1995}), \bibinfo{pages}{5--8}.
\newblock


\bibitem[\protect\citeauthoryear{Messick}{Messick}{1995b}]%
        {messick}
\bibfield{author}{\bibinfo{person}{Samuel Messick}.}
  \bibinfo{year}{1995}\natexlab{b}.
\newblock \showarticletitle{Validity of psychological assessment: Validation of
  inferences from persons' responses and performances as scientific inquiry
  into score meaning}.
\newblock \bibinfo{journal}{{\em American psychologist\/}}
  \bibinfo{volume}{50}, \bibinfo{number}{9} (\bibinfo{year}{1995}),
  \bibinfo{pages}{741}.
\newblock


\bibitem[\protect\citeauthoryear{Nickerson}{Nickerson}{1998}]%
        {nickerson1998confirmation}
\bibfield{author}{\bibinfo{person}{Raymond~S Nickerson}.}
  \bibinfo{year}{1998}\natexlab{}.
\newblock \showarticletitle{Confirmation bias: {A} ubiquitous phenomenon in
  many guises}.
\newblock \bibinfo{journal}{{\em Review of general psychology\/}}
  \bibinfo{volume}{2}, \bibinfo{number}{2} (\bibinfo{year}{1998}),
  \bibinfo{pages}{175}.
\newblock


\bibitem[\protect\citeauthoryear{Polderman, Benyamin, De~Leeuw, Sullivan,
  Van~Bochoven, Visscher, and Posthuma}{Polderman et~al\mbox{.}}{2015}]%
        {polderman2015meta}
\bibfield{author}{\bibinfo{person}{Tinca~JC Polderman}, \bibinfo{person}{Beben
  Benyamin}, \bibinfo{person}{Christiaan~A De~Leeuw},
  \bibinfo{person}{Patrick~F Sullivan}, \bibinfo{person}{Arjen Van~Bochoven},
  \bibinfo{person}{Peter~M Visscher}, {and} \bibinfo{person}{Danielle
  Posthuma}.} \bibinfo{year}{2015}\natexlab{}.
\newblock \showarticletitle{Meta-analysis of the heritability of human traits
  based on fifty years of twin studies}.
\newblock \bibinfo{journal}{{\em Nature genetics\/}} \bibinfo{volume}{47},
  \bibinfo{number}{7} (\bibinfo{year}{2015}), \bibinfo{pages}{702--709}.
\newblock


\bibitem[\protect\citeauthoryear{Pyszczynski and Greenberg}{Pyszczynski and
  Greenberg}{1987}]%
        {pyszczynski1987toward}
\bibfield{author}{\bibinfo{person}{Tom Pyszczynski} {and} \bibinfo{person}{Jeff
  Greenberg}.} \bibinfo{year}{1987}\natexlab{}.
\newblock \showarticletitle{Toward an integration of cognitive and motivational
  perspectives on social inference: {A} biased hypothesis-testing model}.
\newblock \bibinfo{journal}{{\em Advances in experimental social psychology\/}}
   \bibinfo{volume}{20} (\bibinfo{year}{1987}), \bibinfo{pages}{297--340}.
\newblock


\bibitem[\protect\citeauthoryear{Runeson and H{\"o}st}{Runeson and
  H{\"o}st}{2009}]%
        {runeson}
\bibfield{author}{\bibinfo{person}{P. Runeson} {and} \bibinfo{person}{M.
  H{\"o}st}.} \bibinfo{year}{2009}\natexlab{}.
\newblock \showarticletitle{Guidelines for conducting and reporting case study
  research in software engineering}.
\newblock \bibinfo{journal}{{\em Empirical Software Engineering\/}}
  \bibinfo{volume}{14}, \bibinfo{number}{2} (\bibinfo{year}{2009}),
  \bibinfo{pages}{131--164}.
\newblock


\bibitem[\protect\citeauthoryear{Wohlin, Runeson, H{\"o}st, Ohlsson, Regnell,
  and Wessl{\'e}n}{Wohlin et~al\mbox{.}}{2012}]%
        {expbookwholin}
\bibfield{author}{\bibinfo{person}{Claes Wohlin}, \bibinfo{person}{Per
  Runeson}, \bibinfo{person}{Martin H{\"o}st}, \bibinfo{person}{Magnus~C.
  Ohlsson}, \bibinfo{person}{Bj{\"o}rn Regnell}, {and} \bibinfo{person}{Anders
  Wessl{\'e}n}.} \bibinfo{year}{2012}\natexlab{}.
\newblock \bibinfo{booktitle}{{\em Experimentation in software engineering}}.
\newblock \bibinfo{publisher}{Springer}, \bibinfo{address}{Berlin}.
\newblock


\end{thebibliography}

\end{document}